\documentclass[sigconf]{acmart}

\copyrightyear{2025}
\acmYear{2025}
\setcopyright{acmlicensed}\acmConference[KDD '25]{Proceedings of the 31st ACM SIGKDD Conference on Knowledge Discovery and Data Mining V.2}{August 3--7, 2025}{Toronto, ON, Canada}
\acmBooktitle{Proceedings of the 31st ACM SIGKDD Conference on Knowledge Discovery and Data Mining V.2 (KDD '25), August 3--7, 2025, Toronto, ON, Canada}
\acmDOI{10.1145/3711896.3737108}
\acmISBN{979-8-4007-1454-2/2025/08}

\usepackage{xspace}
\usepackage[capitalize,noabbrev,nameinlink]{cleveref}
\usepackage{hyperref}  
\usepackage{diagbox}
\usepackage{multirow}
\usepackage{multicol}
\usepackage{enumitem}
\usepackage{pifont}
\usepackage{amsmath}
\usepackage{colortbl} 
\usepackage{xcolor} 
\DeclareMathOperator*{\argmax}{arg\,max}
\DeclareMathOperator*{\argmin}{arg\,min}

\begin{document}

\title{
Revisiting Self-attention for Cross-domain Sequential Recommendation
}

\author{Clark Mingxuan Ju}
\email{mju@snap.com}
\affiliation{%
  \institution{Snap Inc.}
  \city{Bellevue}
  \state{Washington}
  \country{USA}
}
\author{Leonardo Neves}
\email{lneves@snap.com}
\affiliation{%
  \institution{Snap Inc.}
  \city{Santa Monica}
  \state{California}
  \country{USA}
}
\author{Bhuvesh Kumar}
\email{bkumar4@snap.com}
\affiliation{%
  \institution{Snap Inc.}
  \city{Bellevue}
  \state{Washington}
  \country{USA}
}
\author{Liam Collins}
\email{lcollins2@snap.com}
\affiliation{%
  \institution{Snap Inc.}
  \city{Bellevue}
  \state{Washington}
  \country{USA}
}
\author{Tong Zhao}
\email{tong@snap.com}
\affiliation{%
  \institution{Snap Inc.}
  \city{Bellevue}
  \state{Washington}
  \country{USA}
}
\author{Yuwei Qiu}
\email{yqiu@snap.com}
\affiliation{%
  \institution{Snap Inc.}
  \city{Palo Alto}
  \state{California}
  \country{USA}
}
\author{Qing Dou}
\email{qdou@snap.com}
\affiliation{%
  \institution{Snap Inc.}
  \city{Palo Alto}
  \state{California}
  \country{USA}
}
\author{Sohail Nizam}
\email{snizam@snap.com}
\affiliation{%
  \institution{Snap Inc.}
  \city{Santa Monica}
  \state{California}
  \country{USA}
}
\author{Sen Yang}
\email{syang3@snap.com}
\affiliation{%
  \institution{Snap Inc.}
  \city{Santa Monica}
  \state{California}
  \country{USA}
}
\author{Neil Shah}
\email{nshah@snap.com}
\affiliation{%
  \institution{Snap Inc.}
  \city{Bellevue}
  \state{Washington}
  \country{USA}
}

\renewcommand{\shortauthors}{Clark Mingxuan Ju, et al.}

\newcommand{\method}{\textsc{AutoCDSR}\xspace}
\newcommand{\methodib}{\textsc{AutoCDSR$^+$}\xspace}

\newcommand{\ns}[1]{{\color{purple}[NS: {#1}]}} 
\newcommand{\tz}[1]{{\color{cyan}[TZ: {#1}]}} 

\newcommand{\lc}[1]{{\color{blue}[LC: {#1}]}} 

\newcommand{\bk}[1]{{\color{brown}[BK: {#1}]}} 
\settopmatter{printacmref=false} 
\begin{abstract} 
  Sequential recommendation is a popular paradigm in modern recommender systems.  
  In particular, one challenging problem in this space is \emph{cross-domain sequential recommendation} (CDSR), which aims to predict future behaviors given user interactions across multiple domains.
  Existing CDSR frameworks are mostly built on the self-attention transformer and seek to improve by explicitly injecting additional domain-specific components (e.g. domain-aware module blocks).
  While these additional components help, we argue they overlook the core self-attention module already present in the transformer, a naturally powerful tool to learn correlations among behaviors.
  In this work, we aim to improve the CDSR performance for simple models from a novel perspective of enhancing the self-attention. 
  Specifically, we introduce a Pareto-optimal self-attention and formulate the cross-domain learning as a multi-objective problem, where we optimize the recommendation task while dynamically minimizing the cross-domain attention scores. 
  Our approach \textit{automates knowledge transfer in CDSR} (dubbed as \method) -- it not only mitigates negative transfer but also encourages complementary knowledge exchange among auxiliary domains.
  Based on the idea, we further introduce \methodib, a more performant variant with slight additional cost.
  Our proposal is easy to implement and works as a plug-and-play module that can be incorporated into existing transformer-based recommenders. 
  Besides flexibility, it is practical to deploy because it brings little extra computational overheads without heavy hyper-parameter tuning.
  We conduct experiments over both large-scale production recommender data as well as academic benchmarks, where \method consistently enhances the performance of base transformers, enabling simple models to perform on par with state-of-the-art with less overhead (e.g., \textbf{$\sim$4$\times$} faster than state-of-the-art CDSR models). 
  \method on average improves Recall@10 for SASRec and Bert4Rec by \textbf{9.8\%} and \textbf{16.0\%} and NDCG@10 by \textbf{12.0\%} and \textbf{16.7\%}, respectively. 
  Code is available at \url{https://github.com/snap-research/AutoCDSR}.
\end{abstract}




\maketitle

\section{Introduction}
Recommender systems (RecSys) such as product~\citep{wang2021dcn,schafer1999recommender,judoes}, video~\citep{gomez2015netflix,van2013deep,ju2025learning}, and friend recommendation~\citep{sankar2021graph,ying2018graph,kolodner2024robust} are pivotal in personalizing millions of user experiences and enhancing users’ engagement with web systems.
Particularly, sequential RecSys (SR)~\citep{kang2018self,sun2019bert4rec,zhou2020s3,hidasi2015session} have drawn significant attention; they take sequences constituted by users' historical behaviors as input and predict unseen user interests (e.g., merchandise, movies, short-form videos, etc.).
In SR, a branch of research studies cross-domain SR (CDSR)~\citep{ma2019pi,cao2022contrastive,li2023one,park2023cracking,lin2024mixed}, where user behaviors might come from multiple domains (e.g., products from different categories on e-commerce~\citep{hou2024bridging}, ads and content on social media~\citep{gao2022kuairand,zhang2014large}, etc.). CDSR frameworks assume that cross-domain behaviors of a user are complementary and hence improve the recommendation performance. 

Compared with single-domain SR, CDSR is especially challenging for at least two reasons:
\textit{(i) \textbf{context length explosion}}: user behaviors from multiple domains significantly expand the sequence length, requiring the learning model to effectively capture complicated long-term dependencies~\citep{xu2021long,liu2023linrec}, and \textit{(ii) \textbf{negative transfer}}: cross-domain signal sometimes may be imbalanced and cause asymmetric impact to different domains -- cross-domain behaviors not only augment knowledge but also inadvertently introduce noise, which might mislead the learning model~\citep{park2024pacer,zhu2024modeling,park2023cracking}.

These two challenges are usually \textit{handled separately in existing literature}. 
To mitigate the first one, most existing CDSR frameworks by default explore backbone architectures based on the self-attention transformer~\citep{kang2018self,sun2019bert4rec}, owing to its strong capabilities of modeling long sequences that have been well-demonstrated in other fields~\citep{vaswani2017attention,beltagy2020longformer}.
The second challenge has been studied in several existing works~\citep{park2023cracking,zhu2021cross,zhang2017cross,zang2022survey}; they show that naively stitching cross-domain behaviors into a single sequence and training a single-domain SR results in worse recommendation performance, compared to training single-domain models with single-domain data~\citep{park2023cracking,park2024pacer}.
To alleviate the degradation caused by negative transfer, they explicitly introduce domain-specific components, such as reweighing different domains~\citep{park2024pacer} or deriving domain-aware module blocks~\citep{hwang2024multi,zhang2024mdmtrec}.  
While these components are effective, we argue that they overlook a more effective use of the core self-attention module inherent in the transformer, which is a naturally powerful tool to learn fine-grained correlations among heterogeneous behaviors ~\citep{nagrani2021attention,tsai2019multimodal,xu2023multimodal}. 
In this work, we study whether or not self-attention alone can effectively handle these two challenges at the same time.

\begin{figure}
    \centering
    \vspace{-0.15in}
    \includegraphics{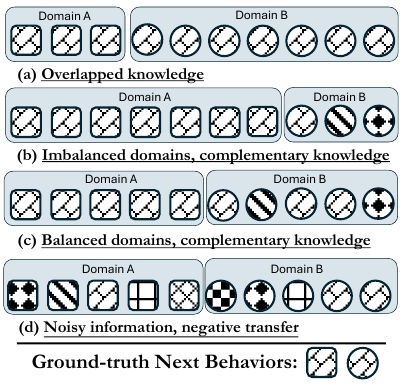}
    \vspace{-0.2in}
    \caption{
    Cross-domain sequences with different characteristics, where tiles indicate behavior semantics.
    \textit{There does not exist a static learning pattern that can well handle knowledge transfer in all scenarios due to asymmetric distribution of domains and inadvertent noises from additional domains}.
    }
    \vspace{-0.1in}
    \label{fig:intro_motivation}
\end{figure}

A naive approach to use self-attention for CDSR is stitching sequences across domains and using a single-domain SR model to perform CDSR.
As an initial study, which will be shown later in \cref{sec:exp}, adopting such an approach can occasionally improve the CDSR performance, relative to existing CDSR benchmarks. (e.g., settings a, b, and c shown in \cref{fig:intro_motivation}). 
However, this approach delivers worse overall recommendation performance compared with training single-domain models with just single-domain data, similar to observations made by existing works~\citep{park2023cracking,park2024pacer}.
Thus, we suspect that these models suffer from \emph{negative transfer} on some examples due to noisy or conflicting information across domains (e.g., \cref{fig:intro_motivation}.d).
To quantitatively validate this understanding, we train a common and simple transformer-based SR model, namely BERT4Rec~\citep{sun2019bert4rec}, on both single-domain and cross-domain sequences, with results shown in \cref{tab:intro_motivtion}.
From this study, we notice that while cross-domain information brings overall performance degradation (i.e., row "Overall" in column (i) vs. that in (ii)), models trained by cross-domain sequences can correctly predict many examples that models trained on a single-domain cannot (i.e., row "Overall" in column (iv) is a lot higher than that in column (i)).
This observation indicates that the self-attention mechanism can successfully learn complementary knowledge in different domains in some cases. 
However, self-attention suffers when cross-domain behaviors are not directly complementary or even contradictory, leading to worse overall performance. 
\begin{table}[t]
\begin{center}
\vspace{-0.05in}
\caption{
The number and percentage of examples that are correctly predicted by BERT4Rec~\citep{sun2019bert4rec} trained on different input sequences for KuaiRand-1K~\citep{gao2022kuairand}. 
Additional settings can be found in \cref{app:intro_setup}.
}
\scalebox{0.9}{
\begin{tabular}{l|c|c||c|c}
\toprule
\multirow{2}{*}{\diagbox{Pred.}{Input}} & \multirow{2}{*}{\vtop{\hbox{\strut 
\ (i)Single}\hbox{\strut \; domain}}} & \multirow{2}{*}{\vtop{\hbox{\strut \ (ii)Cross}\hbox{\strut \; domain}}}  & \multirow{2}{*}{\vtop{\hbox{\strut \ \; (iii)}\hbox{\strut (i) $\cap$ (ii)}}} & \multirow{2}{*}{\vtop{\hbox{\strut \ \; (iv)}\hbox{\strut (i) $\cup$ (ii)}}}  \\
&&&&\\
\midrule
Domain A & \ 52 (18.3\%) & \ 43 (15.1\%) & 28 (10.0\%) & \ 67 (23.6\%)\\
Domain B & 123 (20.4\%) & 111 (18.4\%) & 93 (15.4\%) & 151 (25.1\%)\\
Others   & \ 18 (15.8\%) & \ 22 (19.3\%) & 11 \ (9.6\%) & \ 29 (25.4\%)\\
\midrule
Overall  & 193 (19.3\%) & 176 (17.6\%) & 132 (13.2\%) & 247 (24.7\%)\\
\bottomrule
\end{tabular}
}
\vspace{0.05in}
\label{tab:intro_motivtion}
\end{center}
\vspace{-0.3in}
\end{table}
In light of this study, we ask:
\begin{center}
    \textbf{\textit{Can we enable automatic cross-domain knowledge transfer in CDSR by optimizing self-attention alone?}}
\end{center} 

\noindent Automating knowledge transfer through self-attention offers several transformative advantages: \textit{(i)} it enables seamless integration of CDSR capabilities into any transformer-based single-domain recommender, making cutting-edge cross-domain modeling accessible to widely deployed real-world systems~\citep{fang2024general,pancha2022pinnerformer,de2021transformers4rec}; and \textit{(ii)} it provides a novel perspective on mitigating negative transfer in CDSR by leveraging the \emph{inherent} knowledge-sharing ability of self-attention, paving the way for more effective and adaptive cross-domain recommendations. By harnessing the power of self-attention, we not only enhance the flexibility and scalability of CDSR but also deepen our understanding of how knowledge propagates across domains, opening new directions for future research.

To this end, we propose a novel approach that directly operates on cross-domain attention scores in self-attention by learning a Pareto-optimal self-attention across different domains.
Our method, \method, \textbf{automates knowledge transfer} in CDSR, effectively mitigating negative transfer between conflicting domains while fostering complementary knowledge exchange among auxiliary domains -- \textit{all without introducing additional module blocks}.
We show that that self-attention alone is actually capable of effectively mitigating negative transfer if appropriately optimized.

Specifically, we train a standard single-domain model with two dynamically reconciled objectives. 
The primary objective follows conventional SR training paradigms (e.g., masked token prediction, next-k prediction) to optimize recommendation performance. 
The auxiliary objective minimizes attention scores across behaviors from different domains, selectively regulating cross-domain interactions. 
We frame this as a multi-objective optimization problem, where a Pareto-optimal solution ensures that cross-domain communication occurs only when it benefits the primary recommendation task.
Intuitively, the model prioritizes optimizing recommendation performance with minimal reliance on cross-domain behaviors, increasing cross-domain attention scores only when they provide meaningful improvements.
Building on this idea, we introduce \methodib, a more structured and effective variant of \method. 
In \methodib, we incorporate information bottleneck (IB) tokens~\citep{nagrani2021attention} into multi-domain sequences, restricting cross-domain communication to a dedicated set of Pareto-optimal tokens. 
This structured approach prevents direct information flow across domains, ensuring that knowledge transfer is both selective and efficient, acting as a controlled attention bottleneck.
We summarize our contributions as the following:
\begin{itemize}[leftmargin=*]
    \item 
    In parallel to existing works that mitigate negative transfer with additional model blocks, we seek to automate the knowledge transfer using self-attention that already 
    exists in most sequential recommenders. 
    Our work provides a novel understanding of negative transfer by showing that complementary and conflicting knowledge across domains can be fused and filtered, respectively, via slight changes to self-attention.
    
    \item 
    We propose \method and \methodib, two lightweight plug-and-play mechanisms that improve the CDSR capability for any sequential recommender
    using self-attention.
    \method automates knowledge transfer in CDSR: it not only mitigates negative transfer when behaviors in some domains are noisy, but also encourages complementary knowledge exchange among auxiliary domains. \methodib improves \method by introducing IB tokens to explicitly channel knowledge transfer, still without introducing any additional modules.

    \item 
    We perform extensive experiments on a large-scale production recommender dataset and public benchmarks. \method consistently improves baseline models, allowing simple models to perform comparably to state-of-the-art CSDR methods while reducing overhead (e.g., 4$\times$ faster than state-of-the-art models).
\end{itemize}
\section{Related Work}
\subsection{Sequential Recommendation}
Sequential recommendation aims at predicting user's future behaviors given an ordered list of user's historical interactions. 
Prior to the popularity of transformer models, researchers explored models based on recurrent architectures ~\citep{wu2017recurrent, chung2014empirical} to encode the sequential patterns in user behavior histories, such as GRU4Rec~\citep{hidasi2015session}, STAMP~\citep{liu2018stamp}, NARM~\citep{li2017neural}, etc. 
These works demonstrate that models consuming sequence of user behaviors significantly outperforms pair-wise models such as matrix factorization~\citep{rendle2009bpr}.
After the invention of the transformer~\citep{vaswani2017attention}, sequential recommendation frameworks by default explore backbone model architectures based on this architecture~\citep{kang2018self,sun2019bert4rec}, owing to its strong capabilities of modeling long sequential data that have been well demonstrated in other fields~\citep{vaswani2017attention,beltagy2020longformer}.
For instance, approaches such as SASRec~\citep{kang2018self}, BERT4Rec~\cite{sun2019bert4rec}, SINE~\citep{tan2021sparse}, and LightSANs~\citep{fan2021lighter} train a transformer-based model with supervision signals like causal language modeling or masked language modeling on the user behavior sequence. 
Another branch of research explores textual attributes of behaviors (e.g., reviews and descriptions) and utilizes large language models to conduct sequential recommendation~\citep{zhu2024collaborative,zhao2023survey,wu2024survey,zhang2023recommendation,hou2024large,cui2022m6,geng2022recommendation}.

\subsection{Cross-domain Sequential Recommendation} 
\label{sec:CDSR_relatedworks}
Cross-domain recommendation aims at improving recommendation performance by leveraging information from multiple domains simultaneously.
A branch of early studies explore matrix factorization approaches to model user-item interactions across different domains without considering their sequential nature~\citep{gao2013cross,singh2008relational,liu2020cross,zhu2019dtcdr,li2023one}.
Follow-up research proposes cross-domain sequential recommendation (CDSR) to further improve performance by explicitly injecting additional domain-specific components, such as adding additional supervision signals~\citep{cao2022contrastive}, reweighing different domains~\citep{park2024pacer} and deriving domain-aware module blocks~\citep{hwang2024multi,zhang2024mdmtrec}.
Specifically, $\pi$-net proposes a domain-aware gating mechanism to facilitate knowledge transfer between domains~\citep{ma2019pi}. 
C$^2$DSR leverages graph neural networks that models cross-domain graphs to improve the performance~\citep{cao2022contrastive}.
Similarly, MIFN uses a knowledge graph to enhance CDSR~\citep{ma2022mixed}.
MAN~\citep{lin2024mixed} harnesses additional supervision signals and domain-aware blocks to disentangle information from different domains~\citep{lin2024mixed}.
SyNCRec proposes a cooperative learning framework and utilizes additional domain-specific blocks to advance CDSR~\citep{park2024pacer}.
Although incorporating additional components can be effective, such approaches often overlook the self-attention module in the backbone transformer, which is inherently a powerful tool for capturing fine-grained correlations among heterogeneous behaviors on its own~\citep{nagrani2021attention,tsai2019multimodal,xu2023multimodal}.

\vspace{-0.1in}
\section{Preliminaries}

\textbf{CDSR Problem Formulation}

\noindent This work considers CDSR where user sequences can contain an arbitrary number of domains.
Formally, we denote the set of domains as $\mathcal{D} = \{d_1, d_2, ...\}$ where $|\mathcal{D}| \geq 2$.
Given a domain $d \in \mathcal{D}$, we denote the user sequence in domain $d$ as $X^d = [x^d_1, x^d_2, x^d_3,... ]$, where $x^d_m \in \mathcal{V}^d$ denotes the $m$-th item the user has interacted with in domain $d$ and $\mathcal{V}^d$ refers to all items in domain $d$.
We further denote the cross-domain sequence of a user as $X = [x_1, x_2, x_3,..., x_M]$, where $x_m \in \mathcal{V}$ can come from any domain $d \in \mathcal{D}$, $\mathcal{V}$ denotes the set of all items in all domains, and $M$ refers to the sequence length\footnote{For the ease of notation and reading, here we assume all user sequences have the same length, which is data-dependent and not always true in implementation.}.
$X$ can be generated by stitching $\{X^d\}_{d \in \mathcal{D}}$ together in a way s.t. the user interacted with $x_{m-1}$ earlier than $x_m$ and $|X|=M=\sum_{d \in \mathcal{D}}|X^d|$.
We focus on the retrieval task, where the model maximizes the probability of retrieving the next item that a user will interact with given the current user sequence, formulated as:
\begin{equation}
    \argmax_{x^* \in \mathcal{V}} P(x^* | X, \{X^d\}_{d \in \mathcal{D}}),
    \label{prelim:objective}
\end{equation}
where $x^*$ refers to the next item the user will interact with.

\vspace{0.1in}
\noindent \textbf{Single-domain SR with Self-attention}

\noindent Most state-of-the-art SRs (e.g., SASRec~\citep{kang2018self}, BERT4Rec~\citep{sun2019bert4rec}, PinnerFormer~\citep{pancha2022pinnerformer}, etc) utilize  self-attention~\citep{vaswani2017attention} to model the user sequence. 
Specifically, they usually map individual items in $X$ into learnable vectors, formulated as $\mathbf{H} \in \mathbb{R}^{M \times r} = \text{Emb}(X)$, where $m$-th row (i.e., $\mathbf{h}_m \in \mathbb{R}^r$) represents the vector for item $x_m$, and $\text{Emb}(\cdot)$ refers to the mapping function that converts IDs of a set of items to their corresponding vectors (e.g., embedding table).
Then, multiple layers of self-attention are applied consecutively to $\mathbf{H}$ to facilitate the interaction between items.
At each layer\footnote{Notation for the layer index is neglected for the ease of reading.}, the attention matrix between items is calculated as:
\begin{align}
    & \mathbf{A} = \mathbf{Q}\cdot\mathbf{K}^\intercal, \text{with }\mathbf{Q} = (\mathbf{H}+\mathbf{P})\cdot\mathbf{W}^Q, \mathbf{K} = (\mathbf{H}+\mathbf{P})\cdot\mathbf{W}^K,
\end{align}
where $\mathbf{A}_{i,j}$ is the raw attention score of $j$-th item to $i$-th item\footnote{A causal mask (i.e., $\forall j>i, \mathbf{A}_{i,j}=-\text{inf} $) might be applied per supervision signals.}, $\mathbf{W}^Q \in \mathbb{R}^{r \times r}$ and $\mathbf{W}^K \in \mathbb{R}^{r \times r}$ refer to learnable transformation matrices that convert the original embeddings to queries $\mathbf{Q}$ and keys $\mathbf{K}$ respectively, and $\mathbf{P} \in \mathbb{R}^{M \times r}$ refers to the positional encoding s.t. the model is aware of the ordering of items.
With the attention matrix $\mathbf{A}$, the representation of items in $X$ can be derived as:
\begin{align}
    & \mathbf{H}^* = \text{LN}\Big(\text{FFN}\big(\text{softmax}(\mathbf{A})\cdot\mathbf{V}\big)+\mathbf{H}\Big), \text{with }\mathbf{V}= (\mathbf{H}+\mathbf{P})\cdot\mathbf{W}^V,
    \label{eq:selfattention}
\end{align}
where $\text{LN}(\cdot)$ refers to layer normalization to stabilize the training process, $\text{FFN}(\cdot)$ is stacked feed-forward layers with non-linear transformations, and $\mathbf{W}^V \in \mathbb{R}^{r \times r}$ refers to the transformation matrix that converts the original embedding to values $\mathbf{V}$.
After the derivation of $\mathbf{H}^*$, usually a readout function is used to convert representations of all tokens $\mathbf{H}^*$ into a single user embedding vector $\mathbf{h}^* \in \mathbb{R}^{r}$ that highly correlates with the user's future behaviors (e.g., the hidden representation of the last token as used in SASRec~\citep{kang2018self}, that of the last masked token as used in BERT4Rec~\citep{sun2019bert4rec}, etc). 
To facilitate the training for \cref{prelim:objective}, the model can be optimized by minimizing the cross-entropy loss, formulated as:
\begin{align}
    \mathcal{L}_{\text{rec}} = -\text{log}{\frac{e^{r(x^*)}}{\sum_{x\in V}e^{r(x)}}} \;\text{with}\; r(x) = \frac{\text{Emb}(x)\cdot\mathbf{h}^*}{\left\Vert\text{Emb}(x)\right\Vert \; \left\Vert\mathbf{h}^*\right\Vert} ,
\end{align}
where $x^*$ refers to the ground truth item that the user interacts with next. 
Here we use the cosine similarity to measure affinity between user representation $\mathbf{h}^*$ and item representation $\text{Emb}(x)$. After training, this approach can be used to retrieve the $k$ items with the highest similarity scores for a given user.
\section{Methodology}
\subsection{Motivation}
Existing efforts in CDSR have achieved impressive performance by integrating additional model components~\citep{park2024pacer,park2023cracking,lin2024mixed} or leveraging complex data structures such as graphs~\citep{hou2024ecat,cao2022contrastive}. However, deploying these approaches in online systems presents significant challenges due to their high computational costs, which are further aggravated by large-scale data. Additionally, the simplicity of most industry-standard sequential recommenders~\citep{pancha2022pinnerformer,de2021transformers4rec} creates a substantial gap between cutting-edge research and practical application, making it difficult to adopt advanced models in production environments without significant infrastructural changes.

To bridge this gap, we propose a novel approach that enhances self-attention -- a common component in domain-agnostic SR -- to automate knowledge transfer in CDSR. Unlike prior work that introduces additional module blocks, our method optimizes self-attention directly, leveraging its inherent knowledge-sharing capability to mitigate negative transfer while fostering beneficial cross-domain interactions. This allows seamless integration into existing transformer-based SR with minimal computational overhead, making state-of-the-art cross-domain modeling more accessible to real-world systems. By demonstrating that self-attention alone, if appropriately optimized, is sufficient to address negative transfer, our work offers a scalable and practical solution that unifies research advancements with deployment constraints.
\begin{figure}
    \centering
    \vspace{-0.5in}
    \includegraphics[width=0.48\textwidth]{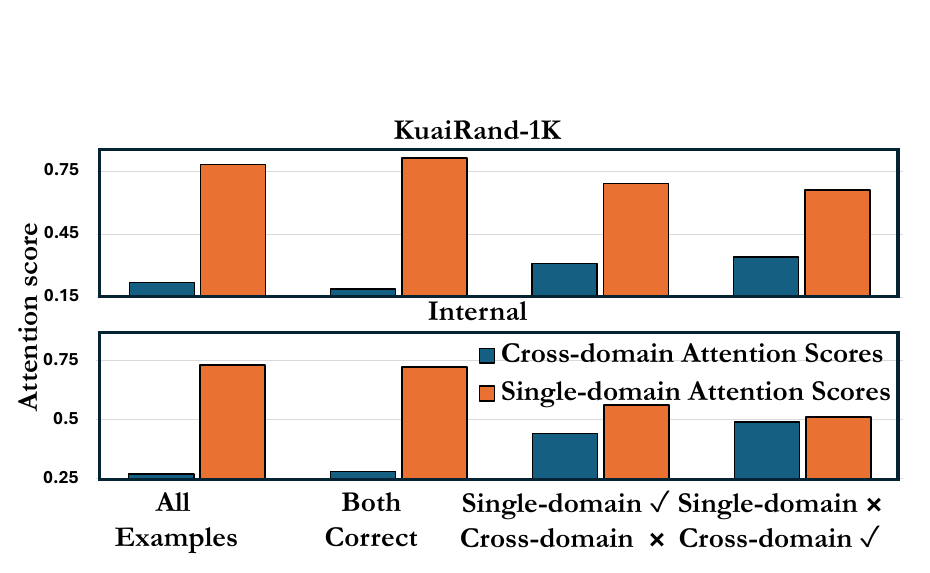}
    \vspace{-0.4in}
    \caption{
    Cross-domain and single-domain attention scores (averaged on all layers and heads) of BERT4Rec trained with cross-domain sequences on two datasets. 
    Negative transfer happens when the model attends too much to unnecessary cross-domain information when single-domain knowledge is sufficient (i.e., the 3rd group).
    }
    \label{fig:attention_score}
    \vspace{-0.2in}
\end{figure}
\subsection{Quantifying Cross-domain Knowledge Transfer via Attention Scores}
\label{sec:attention_score}
We focus on understanding how self-attention behaves in the cross-domain setting and why single-domain recommenders make more mistakes when trained on cross-domain sequences (e.g., \cref{tab:intro_motivtion}).
To begin with, we first quantify the amount of cross-domain communications in sequential recommenders by the values of attention scores between cross-domain items (i.e., $ a_{\text{cd}}$), formulated as:
\begin{equation}
    a_{\text{cd}} = \sum_{i=1}^M \sum_{j=1}^M \text{softmax}(\mathbf{A})_{i,j} \cdot \mathbb{I}(d(x_i) \neq d(x_j)),
\end{equation}
where $d(\cdot)$ returns the domain of an item and $\mathbb{I}(\cdot)$ is an indicator function that returns 1 if the enclosed statement is true otherwise 0.
Intuitively, $a_{\text{cd}}$ quantifies the extent of information exchange between items from different domains. Higher values of $a_{\text{cd}}$ indicate substantial cross-domain interactions, implying that the model is leveraging relationships across diverse domains.
This metric helps evaluate how well the self-attention captures cross-domain signals and highlights potential areas where negative transfer might occur, thereby guiding efforts to optimize the model for enhanced CDSR performance.
Without loss of generality, taking BERT4Rec as an example, we analyze $a_{\text{cd}}$ in BERT4Rec trained on cross-domain sequences and uncover two key insights: 

\begin{itemize}[leftmargin=*]
    \item \textbf{When self-attention fails}: For samples where a single-domain model accurately predicts outcomes (i.e., the `Both Correct' and `Single-domain $\checkmark$, Cross-domain \ding{55}' columns in \cref{fig:attention_score}), attention scores within the same domain dominate those across domains. 
\textit{In these cases, the cross-domain model underperforms when it overemphasizes cross-domain information that is unnecessary for accurate predictions} (i.e., cross-domain score in the 3rd group is higher than that in the 2nd group).
    \item \textbf{When self-attention succeeds}: 
For instances where the single-domain model fails but the cross-domain model succeeds (i.e., the `Single-domain \ding{55}, Cross-domain $\checkmark$' columns in \cref{fig:attention_score}), cross-domain attention scores are more aligned with single-domain scores. 
Here, the cross-domain model effectively attends to relevant information from other domains, leading to improved predictions for these examples.
\end{itemize}

\noindent We conclude that the cross-domain score is a reasonable proxy for measuring knowledge transfer. 
While self-attention effectively facilitates positive knowledge transfer when beneficial in some cases,
it struggles to mitigate negative transfer between unrelated or conflicting domains.
Hence, we seek to improve the CDSR performance for the vanilla self-attention mechanism, such that it encourages cross-domain knowledge transfer when needed (i.e., the `Single-domain \ding{55}, Cross-domain $\checkmark$' case in \cref{fig:attention_score}) and suppressed when redundant (i.e., the `Single-domain $\checkmark$, Cross-domain \ding{55}' case). 

\subsection{Proposed Method: \method}

Based on the observation above, we propose to mitigate negative transfer in CDSR by directly minimizing the cross-domain attention scores $a_{cd}$. 
We initially formalize the corresponding loss as:
\begin{equation}
    \mathcal{L} = \alpha_1 \cdot \mathcal{L}_{\text{rec}} + \alpha_2 \cdot \mathcal{L}_{\text{cd-attn}}, \ \text{with} \ \mathcal{L}_{\text{cd-attn}} = a_\text{cd},
    \label{eq:old_loss}
\end{equation}
where $\alpha_1, \alpha_2$ are coefficients that control the strength of suppressing the knowledge transfer between domains.
One can manually tune the $\alpha$ values according to dataset-level prior knowledge (e.g., assigning a higher $\alpha_2$ value for datasets where cross-domain information is less complementary, or vice versa).
However, even within the same dataset, different user behavior sequences might require different levels of cross-domain attention, as the relevance of cross-domain information can vary depending on the specific context, user preferences, or item interactions within each sequence.
As we have shown later in this paper (i.e., \cref{fig:hyper_param}), manually tuning $\alpha$ usually does not lead to the optimal performance. 
Furthermore, manually tuning $\alpha$ is not feasible per sequence, calling for an automatic tuning of $\alpha$ in a more fine-grained manner. 
An overview figure of our proposal (\cref{fig:overview}) can be found in \cref{sec:app}.
\subsubsection{\textbf{Two-task Reconciliation that Promotes Preference-aware Pareto Optimality}}
\ \\
In order to automate the knowledge transfer between domains, we re-formulate the optimization problem in \cref{eq:old_loss} as a multi-task learning (MTL) problem with two tasks:
\begin{equation}
    \min_{\boldsymbol{\theta}} \boldsymbol{\mathcal{L}}(\boldsymbol{\theta}) = \min_{\boldsymbol{\theta}} \Big(\mathcal{L}_{\text{rec}}(\boldsymbol{\theta}), \mathcal{L}_{\text{cd-attn}}(\boldsymbol{\theta})\Big),
\end{equation} where $\boldsymbol{\theta}$ refers to the set of model parameters.
We reconcile between the recommendation loss $\mathcal{L}_{\text{rec}}$ and the cross-domain attention loss $\mathcal{L}_{\text{cd-attn}}$, \emph{ensuring that the self-attention increases cross-domain scores only when cross-domain information improves the recommendation task}.
To achieve this goal, we adopt ideas from the multi-task learning community~\citep{sener2018multi,lin2019pareto,lee2016asymmetric} by learning a preference-aware Pareto-optimal solution. 
Formally, Pareto optimality is defined as:
\begin{definition}[Pareto Optimality] 

For an MTL problem with $k$ tasks,
a solution $\boldsymbol{\theta}^\star$ is Pareto-optimal iff there does not exist another solution that dominates $\boldsymbol{\theta}^\star$. $\boldsymbol{\theta}^\star$ dominates $\boldsymbol{\hat{\theta}}$ if for every task $k$, we have $\mathcal{L}_k(\boldsymbol{\theta}^\star) \leq \mathcal{L}_k(\boldsymbol{\hat{\theta}})$, 
and for some task $k'$, $\mathcal{L}_{k'}(\boldsymbol{\theta}^\star) <\mathcal{L}_{k'}(\boldsymbol{\hat{\theta}})$.
\end{definition}
\noindent Intuitively, if a model is Pareto-optimal, it is impossible to further lower the error on any loss without hurting at least one other. 
Seeking a Pareto-optimal solution would not be necessary if the CDSR model could perfectly minimize both losses independently (i.e., achieve optimal recommendation performance without leveraging any cross-domain information). However, this contradicts the premise that cross-domain information provides complementary knowledge and can enhance CDSR. Even in scenarios where cross-domain signals offer limited benefits, our approach ensures that knowledge transfer is adaptively regulated -- preventing negative transfer while preserving the performance of each domain.

The Pareto front\footnote{The set of all Pareto optimal solutions with different preferences over tasks.}
contains many Pareto optimal solutions.
In order to derive a Pareto optimal solution that satisfies our needs, we need the Pareto optimal model to fall in the subregion favoring the recommendation task.
We start from learning a Pareto optimal solution without any preference. 
We explore Multiple Gradient Descent Algorithm~\citep{desideri2012multiple} to derive a Pareto optimal solution agnostic of the preference between $\mathcal{L}_{\text{rec}}$ and $\mathcal{L}_{\text{cd-attn}}$. 
In this algorithm, a model theoretically converges to Pareto optimality if we utilize the gradient that satisfies the saddle-point test~\citep{sener2018multi}. 
Finding such a gradient is equivalent to finding the direction with minimum norm in the convex hull\footnote{The surface constituted by all linear weighted scalarization of descent directions.} constructed by the descent directions of our two tasks, formulated as: 
\begin{align}
\begin{split}
     \min_{\alpha_1, \alpha_2} &\Big\lvert\Big\lvert \alpha_1 \cdot \nabla_{\boldsymbol{\theta}} \mathcal{L}_{\text{rec}}(\boldsymbol{\theta}) + \alpha_2 \cdot \nabla_{\boldsymbol{\theta}} \mathcal{L}_{\text{cd-attn}}(\boldsymbol{\theta})\Big\rvert\Big\lvert_F, \\
    & \;\;\;\;\text{s.t.} \;\; \alpha_1 + \alpha_2 = 1 \;\;\text{and} \;\; \alpha_1, \alpha_2 \geq 0,
    \label{eq:opt}
    \end{split}
\end{align}
where $\nabla_{\boldsymbol{\theta}} \mathcal{L}_{\text{rec}}(\boldsymbol{\theta}) \in \mathbb{R}^{|\boldsymbol{\theta}|}$ refers to the gradient of 
$\boldsymbol{\theta}$ for $\mathcal{L}_{\text{rec}}$ and $\lvert\lvert \cdot \rvert\lvert_F$ is Frobenius norm. \
Finding a solution to \cref{eq:opt} is rather straight-forward: if the norm of one gradient direction is smaller than the inner product between both directions, the solution is the gradient with the smaller norm (i.e., ($\alpha_1=0, \alpha_2=1$) or vice versa).
Otherwise, $\alpha_1$ can be derived by calculating the vector perpendicular to the convex hull, as follows:
\begin{equation}
    \alpha_1 = \frac{\nabla_{\boldsymbol{\theta}} \mathcal{L}_{\text{cd-attn}}(\boldsymbol{\theta}) \cdot \Big( \nabla_{\boldsymbol{\theta}} \mathcal{L}_{\text{cd-attn}}(\boldsymbol{\theta})- \nabla_{\boldsymbol{\theta}} \mathcal{L}_{\text{rec}}(\boldsymbol{\theta})\Big)^\intercal}{\Big\rvert\Big\lvert \nabla_{\boldsymbol{\theta}} \mathcal{L}_{\text{cd-attn}}(\boldsymbol{\theta}) - \nabla_{\boldsymbol{\theta}} \mathcal{L}_{\text{rec}}(\boldsymbol{\theta})\Big\rvert\Big\lvert_F}.
    \label{eq:two_task}
\end{equation}
While training the model with \cref{eq:two_task} theoretically guarantees Pareto optimality, as proven by previous works~\citep{desideri2012multiple}, the final solution falls near a region in the Pareto front where errors on different tasks are balanced~\citep{ju2022multi}, which cannot be directly applied to our scenario.

To solve this challenge, inspired by existing works in the multi-task learning community~\citep{lin2019pareto}, we propose to partition the Pareto front into multiple sub-regions and enforce the model to converge to the preferred sub-region rather than a random point in the entire space.
Specifically, to partition the Pareto front into $K$ sub-regions, we first define $K+1$ preference vectors, denoted as:
$
    \big\{\mathbf{p}_k \in \mathbb{R}_+^2= \big(\cos{(\frac{k\pi}{2K})}, \sin{(\frac{k\pi}{2K})}\big) \big\rvert k=0,\ldots,K \big\},
$
where $\mathbf{p}_0 = (1, 0)$ and $\mathbf{p}_K = (0, 1)$ refer to preference vectors totally biased to the main recommendation task and cross-domain attention respectively, and other vectors are in-between preferences. 

With $K+1$ preference vectors, we select $\mathbf{p}_1$ as our preference vector to prioritize $\mathcal{L}_{\text{rec}}(\boldsymbol{\theta})$ over $\mathcal{L}_{\text{cd-attn}}(\boldsymbol{\theta})$ yet still allow cross-domain knowledge transfer. 
Intuitively, such a Pareto optimal solution enforces the self-attention to only increase cross-domain attention scores when the recommendation task can be improved (i.e., increasing $\mathcal{L}_{\text{cd-attn}}(\boldsymbol{\theta})$ can accordingly offset the decrease in $\mathcal{L}_{\text{rec}}(\boldsymbol{\theta})$). 
To obtain a preference-aware Pareto optimality, given a loss vector $\boldsymbol{\mathcal{L}}(\boldsymbol{\theta})$, we identify the set of preference vectors $\mathcal{S}$ that $\boldsymbol{\mathcal{L}}(\boldsymbol{\theta})$ better aligns with than $\mathbf{p}_1$, formulated as:
\begin{equation}
    \mathcal{S} = \big\{\mathbf{p}_k \big\rvert \big(\mathbf{p}_k^\intercal \cdot \boldsymbol{\mathcal{L}}(\boldsymbol{\theta})- \mathbf{p}_1^\intercal \cdot \boldsymbol{\mathcal{L}}(\boldsymbol{\theta})\big)\geq0\big\}.
\end{equation}
Each $\mathbf{p}_k \in \mathcal{S}$ is an active constraint w.r.t. $\mathbf{p}_1$, 
indicating that the current model overly attends to cross-domain information yet not improving the recommendation task as much.
In this scenario, we need to increase $\alpha_1$ to mitigate the cross-domain negative transfer. 
To incorporate $\mathcal{S}$, we treat active constraints as additional tasks in \cref{eq:opt} and reformulate it as:
\begin{align}
\begin{split}
     \min_{\boldsymbol{\beta}} &\Big\lvert\Big\lvert \sum_{\mathbf{p}_k \in \mathcal{S}} \beta_k \cdot \nabla_{\boldsymbol{\theta}} \big(\mathbf{p}_k^\intercal \cdot \boldsymbol{\mathcal{L}}(\boldsymbol{\theta})- \mathbf{p}_1^\intercal \cdot \boldsymbol{\mathcal{L}}(\boldsymbol{\theta})\big) \Big\rvert\Big\lvert_F, \\
    & \text{s.t.} \;\; \sum_{\mathbf{p}_k \in \mathcal{S}} \beta_k = 1 \;\;\text{and} \;\; \forall{k} \; \beta_k \geq 0.
    \label{eq:opt_new}
    \end{split}
\end{align}
We can solve \cref{eq:opt_new} by extending \cref{eq:two_task}.
Utilizing Frank-Wolfe algorithm~\citep{jaggi2013revisiting}, we approximate \cref{eq:opt_new} with multiple iterations, each of which is a special case of \cref{eq:two_task}. 
We first assign each task the same weight (i.e., $\boldsymbol{\beta} \in \mathbb{R}_+^{|\mathcal{S}|} = \frac{1}{|\mathcal{S}|}$) and at each iteration find the single task that correlated with other tasks the least given the current weight assignment. 
Then we treat the least correlated task as one task and the combination of remaining tasks as the other to update the weight as the following:
\begin{align}
    \begin{split}
        & \boldsymbol{\beta} := (1-\eta)\cdot\boldsymbol{\beta} + \eta\cdot\mathbf{e}_t,
         \; \text{with}\; \eta = \frac{\nabla_{\boldsymbol{\theta}}'  \cdot \Big( \nabla_{\boldsymbol{\theta}}'- \nabla_{\boldsymbol{\theta}}^t\Big)^\intercal}{\Big\rvert\Big\lvert \nabla_{\boldsymbol{\theta}}' - \nabla_{\boldsymbol{\theta}}^t \Big\rvert\Big\lvert_F},\\
         & \nabla_{\boldsymbol{\theta}}' = \sum_{\mathbf{p}_k \in \mathcal{S}} \beta_k \cdot \nabla_{\boldsymbol{\theta}} \big(\mathbf{p}_k^\intercal \cdot \boldsymbol{\mathcal{L}}(\boldsymbol{\theta})- \mathbf{p}_1^\intercal \cdot \boldsymbol{\mathcal{L}}(\boldsymbol{\theta})\big),\\
         & t = \argmin_{t} \nabla_{\boldsymbol{\theta}}' \cdot \nabla_{\boldsymbol{\theta}} \big(\mathbf{p}_t^\intercal \cdot \boldsymbol{\mathcal{L}}(\boldsymbol{\theta})- \mathbf{p}_1^\intercal \cdot \boldsymbol{\mathcal{L}}(\boldsymbol{\theta})\big), 
    \end{split}
    \label{eq:pareto}
\end{align}
where $\mathbf{e}_t \in \{0,1\}^{|\mathcal{S}|}$ refers to the one-hot vector with $t$-th element equals to 1.
The procedure described above iterates until $\eta$ becomes smaller than a pre-defined threshold or the procedure reaches a fixed number of iterations.
As proven in the previous work~\citep{ju2022multi}, optimizing the model with \cref{eq:pareto} and theoretically deliver well-approximated solutions within reasonable iterations (e.g., 100). 
We include sensitivity studies in the \cref{app:sensitivity} 
and \method converges within reasonable numbers of iterations. 

\subsubsection{\textbf{\methodib: Further enhancement with Pareto-optimal Information Bottleneck Tokens}}
\ \\
Research in various domains has explored the use of placeholder tokens in input sequences to facilitate knowledge transfer~\citep{nagrani2021attention,tsai2019multimodal}. These tokens serve as information bottlenecks (IB), constraining the transfer process to occur exclusively through them.
Inspired by this general principle, we introduce \methodib, which integrates IB tokens into multi-domain sequences and restricts cross-domain communication to these tokens. Unlike \method, which permits direct information flow across domains, \methodib enforces structured information transfer by designating a small set of dedicated Pareto-optimal tokens as attention bottlenecks.
In our approach, individual single-domain sequences, along with IB tokens, are processed separately. IB tokens attend to elements within their respective domains to capture domain-specific knowledge, while cross-domain information exchange occurs exclusively between IB tokens associated with different domains. Following this exchange, items within each domain re-attend to the updated IB tokens to acquire transferred knowledge from other domains.
Similar to \method, \methodib employs Pareto-optimal cross-domain self-attention, with cross-domain attention scores computed as:
\begin{equation}
    a_{\text{cd}} = \sum_{d\in\mathcal{D}}\sum_{i=T}^{M^d} \sum_{j=1}^{T} \mathbf{A}_{i,j}^d,
    \label{eq:ib_attn}
\end{equation}
where $T$ refers to the number of injected IB tokens and $M^d$ refers to the sequence length for domain $d$.
Following the formulation in \cref{eq:selfattention}, the cross-domain information exchange can be achieve by mixing representations of IB tokens in different domains:
\begin{equation}
    \mathbf{H}^* = \text{Combine}(\{\mathbf{H}^{d*}[1:T]|d\in\mathcal{D}\}),
\end{equation}
where $\text{Combine}(\cdot)$ can be any combination function to aggregate different domains\footnote{We use a simple element-wise summation for the combination method. Advanced combination strategies can be helpful but out-of-scope for this research.} and $\mathbf{H}^{d*}$ refers to the final sequence embedding in domain $d$.
We can train \methodib using the same optimization we propose in \cref{eq:opt_new} with attention scores derived in \cref{eq:ib_attn}.
\methodib offers a more structured flow of information that not only facilitates effective knowledge extraction within each domain but also ensures that cross-domain transfer occurs in a controlled manner.
Compared with \method, \methodib further reduces the risk of information dilution or conflicts, with the cost of slightly additional computational overheads and model complexity.

\begin{table*}[h]
\vspace{-0.1in}
\caption{The performance (Recall@5/Recall@10/NDCG@5/NDCG@10) of single-domain as well as cross-domain models on the Amazon-Review dataset. We explore the same setting adopted in the previous work~\citep{park2024pacer}. 
Perf. $\uparrow$ refers to the percentage of improvement brought by variants of \method. 
}
\resizebox{2.\columnwidth}{!}{
    \begin{tabular}{l|cccc|cccc|cccc|cccc|cccc}
    \toprule
    \multirow{2}{*}{Method} & \multicolumn{20}{c}{Domain (Recall@5/Recall@10/NDCG@5/NDCG@10)} \\
    \cmidrule(r){2-21}
    &\multicolumn{4}{c}{Book}&\multicolumn{4}{c}{Clothing}&\multicolumn{4}{c}{Video}&\multicolumn{4}{c}{Toy}&\multicolumn{4}{c}{Sports}\\
    \midrule
    \multicolumn{21}{c}{Single-domain Models with single-domain Input Sequences} \\
    \midrule
    GRU4Rec & 0.265 & 0.348 & 0.195 & 0.222 & 0.334 & 0.434 & 0.242 & 0.274 & 0.344 & 0.451 & 0.242 & 0.277 & 0.356 & 0.464 & 0.256 & 0.291 & 0.551 & 0.661 & 0.416 & 0.452\\
    SASRec & 0.261 & 0.355 & 0.187 & 0.218 & 0.306 & 0.415 & 0.219 & 0.255 & 0.345 & 0.473 & 0.243 & 0.284 & 0.352 & 0.476 & 0.251 & 0.291 & 0.623 & 0.744 & 0.474 & 0.514\\
    BERT4Rec & 0.208 & 0.296 & 0.144 & 0.172 & 0.277 & 0.387 & 0.192 & 0.228 & 0.311 & 0.433 & 0.214 & 0.254 & 0.320 & 0.442 & 0.225 & 0.264 & 0.549 & 0.676 & 0.415 & 0.456\\
    \midrule
    \multicolumn{21}{c}{Cross-domain Models with Cross-domain Input Sequences} \\
    \midrule 
    SyNCRec & 0.337 & 0.433 & 0.249 & 0.280 & 0.366 & 0.480 & 0.262 & 0.298 & 0.442 & 0.567 & 0.320 & 0.361 & 0.438 & 0.553 & 0.317 & 0.355 & 0.724 & 0.810 & 0.597 & 0.625\\
    C$^2$DSR & 0.320 & 0.420 & 0.241 & 0.271 & 0.234 & 0.345 & 0.182 &  0.218 & 0.265 & 0.379 & 0.184 & 0.221 & 0.256 & 0.367 & 0.182 & 0.218 & 0.362 & 0.501 & 0.249 & 0.294\\
    CGRec & 0.261 & 0.352 & 0.189 & 0.218 & 0.339 & 0.452 & 0.242 & 0.278 & 0.384 & 0.504 & 0.274 & 0.313 & 0.394 & 0.512 & 0.284 & 0.323 & 0.655 & 0.758 & 0.519 & 0.553\\
    \midrule 
    \multicolumn{21}{c}{Single-domain Models with Cross-domain Input Sequences} \\
    \midrule
    SASRec$_{\text{cd}}$ & 0.256 & 0.348 & 0.175 & 0.203 & 0.293 & 0.395 & 0.220 & 0.256 & 0.353 & 0.485 & 0.234 & 0.272 & 0.330 & 0.443 & 0.236 & 0.276 & 0.605 & 0.722 & 0.467 & 0.506\\
    \; +\method & 0.328 & 0.423 & 0.242 & 0.277 & 0.368 & 0.467 & 0.255 & 0.297 & 0.439 & 0.556 & 0.314 & 0.355 & 0.421 & 0.530 & 0.313 & 0.351 & 0.719 & 0.805 & 0.572 & 0.601\\
    \rowcolor[gray]{0.9} Perf. $\uparrow$ (\%) & 27.9\% & 21.6\% & 38.0\% & 36.6\% & 25.4\% & 18.3\% & 15.7\% & 15.9\% & 24.4\% & 14.7\% & 34.1\% & 30.6\% & 27.5\% & 19.5\% & 32.6\% & 27.3\% & 18.8\% & 11.4\% & 22.4\% & 18.8\%\\ 
    \; +\methodib & 0.340 & 0.436 & 0.250 & 0.274 & 0.356 & 0.471 & 0.262 & 0.292 & 0.429 & 0.560 & 0.322 & 0.361 & 0.436 & 0.541 & 0.309 & 0.352 & 0.708 & 0.792 & 0.590 & 0.628\\
    \rowcolor[gray]{0.9} Perf. $\uparrow$ (\%) & 32.6\% & 25.2\% & 42.8\% & 35.2\% & 21.6\% & 19.2\% & 19.1\% & 14.1\% & 21.4\% & 15.5\% & 37.5\% & 32.7\% & 32.1\% & 22.1\% & 30.7\% & 27.3\% & 17.0\% & 9.6\% & 26.2\% & 24.1\%\\ 
    \midrule 
    BERT4Rec$_{\text{cd}}$ & 0.220 & 0.313 & 0.151 & 0.181 & 0.293 & 0.410 & 0.198 & 0.235 & 0.329 & 0.458 & 0.221 & 0.262 & 0.335 & 0.463 & 0.232 & 0.273 & 0.576 & 0.709 & 0.440 & 0.483\\
    \; +\method & 0.292 & 0.415 & 0.221 & 0.246 & 0.364 & 0.452 & 0.249 & 0.261 & 0.418 & 0.520 & 0.278 & 0.340 & 0.413 & 0.513 & 0.298 & 0.317 & 0.671 & 0.785 & 0.511 & 0.601\\
    \rowcolor[gray]{0.9} Perf. $\uparrow$ (\%) & 32.7\% & 32.5\% & 46.2\% & 35.9\% & 24.3\% & 10.1\% & 25.7\% & 11.1\% & 26.9\% & 13.5\% & 25.8\% & 29.8\% & 23.2\% & 10.7\% & 28.3\% & 16.2\% & 16.5\% & 10.7\% & 16.2\% & 24.4\%\\ 
    \; +\methodib & 0.307 & 0.395 & 0.225 & 0.267 & 0.338 & 0.472 & 0.246 & 0.279 & 0.403 & 0.535 & 0.313 & 0.358 & 0.423 & 0.550 & 0.307 & 0.346 & 0.667 & 0.804 & 0.557 & 0.592\\
    \rowcolor[gray]{0.9} Perf. $\uparrow$ (\%) & 39.5\% & 26.2\% & 48.8\% & 47.6\% & 15.4\% & 15.0\% & 24.4\% & 18.7\% & 22.4\% & 16.8\% & 41.8\% & 36.8\% & 26.2\% & 18.7\% & 32.5\% & 26.6\% & 15.8\% & 13.3\% & 26.6\% & 22.5\%
    \\ 
    \bottomrule
    \end{tabular}
    }
\label{tab:amazon}
\end{table*}
\begin{table*}[h]
\vspace{-0.1in}
\caption{The performance (NDCG@10/NDCG@20/Recall@10/Recall@20) of single-domain as well as cross-domain models on KuaiRand-1K and Internal datasets. Perf. $\uparrow$ refers to the percentage of improvement brought by variants of \method. }
\vspace{-0.1in}
\resizebox{2.\columnwidth}{!}{
    \begin{tabular}{l|cccc|cccc|cccc|cccc}
    \toprule
    \multirow{3}{*}{Method} & \multicolumn{16}{c}{Domain (NDCG@10/NDCG@20/Recall@10/Recall@20)} \\
    \cmidrule(r){2-17}
    & \multicolumn{8}{c}{KuaiRand-1K} & \multicolumn{8}{|c}{Internal} \\
    \cmidrule(r){2-9} \cmidrule(r){10-17}
    &\multicolumn{4}{c}{Type A}&\multicolumn{4}{c}{Type B}&\multicolumn{4}{|c}{Tab A}&\multicolumn{4}{c}{Tab B}\\
    \midrule
    \multicolumn{17}{c}{Single-domain Models with single-domain Input Sequences} \\
    \midrule
    SASRec & 0.019 & 0.030 & 0.044 & 0.098 & 0.031 & 0.039 & 0.059 & 0.133 & 0.053 & 0.066 & 0.100 & 0.152 & 0.228 & 0.247 & 0.361 & 0.454 \\
    BERT4Rec & 0.024 & 0.035 & 0.054 & 0.124 & 0.037 & 0.050 & 0.077 & 0.153 & 0.054 & 0.068 & 0.104 & 0.158 & 0.238 & 0.268 & 0.382 & 0.490 \\
    \midrule
    \multicolumn{17}{c}{Single-domain Models with Cross-domain Input Sequences} \\
    \midrule 
    SASRec$_{\text{cd}}$ & 0.022 & 0.031 & 0.041 & 0.093 & 0.033 & 0.046 & 0.063 & 0.135& 0.053 & 0.066 & 0.101 & 0.154 & 0.217 & 0.241 & 0.364 & 0.461 \\
    \; +\method &  0.028 & 0.039 & 0.049 & 0.106 &  0.037 & 0.051 & 0.069 & 0.141 & 0.054 & 0.068 & 0.108 & 0.165 & 0.230 & 0.255 & 0.377 & 0.474\\
    \rowcolor[gray]{0.9} Perf. $\uparrow$ (\%) & 27.2\% & 25.8\% & 19.5\% & 13.9\% & 12.1\% & 10.8\% & 9.5\% & 4.4\% & 2.7\% & 3.2\% & 6.8\% & 7.1\% & 6.0\% & 5.8\% & 3.4\% & 2.7\%\\
    \; +\methodib &  0.026 & 0.037 & 0.046 & 0.102 &  0.039 & 0.054 & 0.073 & 0.153  & 0.057 & 0.071 & 0.107 & 0.165 & 0.243 & 0.269 & 0.406 & 0.514\\
    \rowcolor[gray]{0.9} Perf. $\uparrow$ (\%) & 18.1\% & 19.3\% & 12.2\% & 9.6\% & 18.1\% & 17.3\% & 15.8\% & 13.3\% & 8.1\% & 7.8\% & 6.3\% & 6.8\% & 11.7\% & 11.7\% & 11.6\% & 11.6\%\\
    \midrule
    BERT4Rec$_{\text{cd}}$ & 0.027 & 0.045 & 0.060 & 0.134 & 0.031 & 0.044 & 0.063 & 0.122 & 0.054 & 0.068 & 0.103 & 0.158 & 0.228 & 0.254 & 0.381 & 0.483 \\
    \; +\method & 0.031 & 0.049 & 0.071 & 0.173 & 0.043 & 0.052 & 0.084 & 0.165 & 0.058 & 0.074 & 0.110 & 0.168 & 0.241 & 0.269 & 0.402 & 0.513\\
    \rowcolor[gray]{0.9} Perf. $\uparrow$ (\%) & 14.7\% & 8.8\% & 18.3\% & 29.1\% & 38.7\% & 18.1\% & 33.3\% & 26.2\% & 7.4\% & 8.8\% & 6.8\% & 6.3\% & 5.7\% & 5.9\% & 5.5\% & 6.2\%\\
    \; +\methodib & 0.033 & 0.049 & 0.066 & 0.139 & 0.037 & 0.064 & 0.103 & 0.178 & 0.059 & 0.077 & 0.113 & 0.170 & 0.254 & 0.281 & 0.415 & 0.522 \\
    \rowcolor[gray]{0.9} Perf. $\uparrow$ (\%) & 22.2\% & 8.8\% & 10.0\% & 3.7\% & 19.3\% & 45.4\% & 58.7\% & 45.9\% & 9.2\% & 13.2\% & 9.7\% & 7.6\% & 11.4\% & 10.6\% & 8.9\% & 8.0\%\\
    \bottomrule
    \end{tabular}
    }
\label{tab:others}
\vspace{-0.1in}
\end{table*}
\vspace{-0.05in}
\section{Experiment}
\label{sec:exp}
We conduct systematic experiments to validate our proposal, leading to three research questions (RQs):
(\textbf{RQ1}) What is the performance boost that \method can bring to simple transformers to improve their cross-domain performance?
(\textbf{RQ2}) Given the performance boost, how much additional computational overhead does \method introduce?
(\textbf{RQ3}) How and why does \method improve the CDSR performance?
\vspace{-0.1in}
\subsection{Experimental Setup}
\noindent \textbf{Datasets}:
We conduct experiments on two academic benchmark datasets (i.e., \textsc{KuaiRand-1K}~\citep{gao2022kuairand} and \textsc{Amazon Reviews}~\citep{mcauley2015image,hou2024bridging}) as well as one industrial dataset collected internally at a leading social media platform spanning interactions over multiple mobile in-app surfaces (dubbed as \textsc{Internal}).
For \textsc{KuaiRand-1K}, we use music type as the domain indicator; for \textsc{Amazon Reviews}, following the same setting explored in the previous work~\citep{park2024pacer}, we include user behaviors from Book, Clothing, Video, Toy, and Sports domains. 
Whereas for \textsc{Internal}, we use the in-app surface where the user behavior comes from as the domain indicator. 

\noindent \textbf{Baselines}:
We compare \method and \methodib against a broad range of baselines, including domain-agnostic sequential recommenders (i.e., GRU4Rec~\citep{hidasi2015session}, SASRec~\citep{kang2018self}, BERT4Rec~\citep{sun2019bert4rec}) and domain-aware sequential recommenders (i.e., SyNCRec~\citep{park2024pacer}, C$^2$DSR~\citep{cao2022contrastive}, and CGRec~\citep{cao2022contrastive}). 
As \method and \methodib improve the self-attention mechanism in the transformer model, we utilize them as plug-and-play modules and apply them to vanilla transformer-based CDSR models (i.e., SASRec and BERT4Rec). 

\noindent \textbf{Evaluation Protocol}:
We evaluate with metrics adopted in previous works, including NDCG and Recall~\citep{park2024pacer,kang2018self}. All studies are repeated for 5 times and mean values are reported.

\noindent Due to space limitation, we include other details in \cref{app:dataset}. 
\vspace{-0.05in}
\subsection{Comparison Experiment}

We analyze the performance of single-domain transformer models enhanced by \method as well as \methodib and compare them against baselines, with results shown in \cref{tab:amazon} and \cref{tab:others}. 
Firstly, by feeding cross-domain sequences into single-domain models without special treatment, most of the time we observe performance degradation (i.e., SASRec$_{\text{cd}}$/Bert4Rec$_{\text{cd}}$ vs. SASRec/Bert4Rec), confirming the negative transfer between different domains when they are naively stitched together (e.g., NDCG@10 dropped from 0.218 to 0.203 for SASRec in the book domain of Amazon-Review, etc).  
\method and \methodib are plug-and-play ideas that can be used unanimously to enhance the cross-domain performance of domain-agnostic models by automating knowledge transfer between domains. 
We apply them to both SASRec$_{\text{cd}}$ and Bert4Rec$_{\text{cd}}$ and observe if our proposal can improve the CDSR performance. 
Across all datasets, adding either \method or \methodib to base models can significantly improve their CDSR performance. 
Specifically, for Amazon-Review dataset, \method significantly improves the CDSR performance over 20\% in 12 out of 20 cases for SASRec, and 14 out of 20 cases for BERT4Rec. 
\textbf{Enhanced by \method and \methodib, simple models (e.g., BERT4Rec$_{\text{cd}}$ + \method) can perform on par with most of state-of-the-art CDSR models or sometimes even outperform some}. 

In some cases, SASRec$_{\text{cd}}$ and BERT4Rec$_{\text{cd}}$ can already improve the performance of base models (e.g., NDCG for Type A in \textsc{KuaiRand-1K}). 
By adding our proposal, we can further amplify these gains while mitigating the negative transfer, validating the capability of \method for automating knowledge transfer.
To answer RQ1, we observe that \method significantly enhances simple transformers, effectively automating knowledge transfer and enhancing their CDSR performance, enabling them to match or even surpass state-of-the-art CDSR models. 
\begin{figure}[t]
    \centering
    \includegraphics[width=0.45\textwidth]{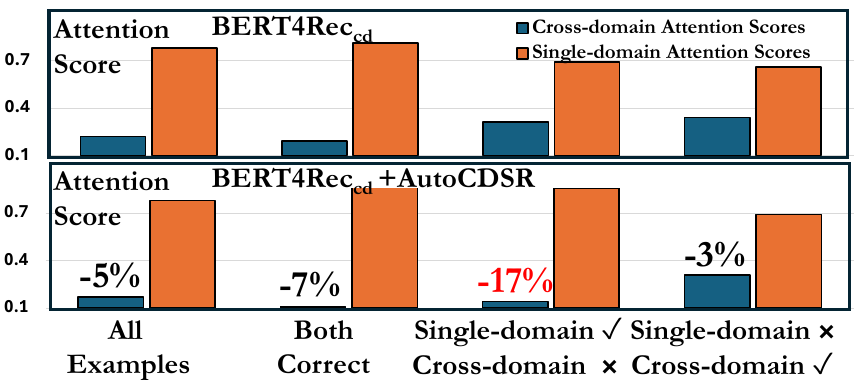}
    \caption{
    The distribution of attention scores across different strata in \textsc{KuaiRand-1K} with the deployment of \method. Cross-domain attention scores for samples suffering from negative transfer are reduced significantly by \method.
    }
    \label{fig:attn_score_exp}
\end{figure}
\begin{figure}[t]
    \centering
    \vspace{-0.1in}
    \includegraphics[width=0.48\textwidth]{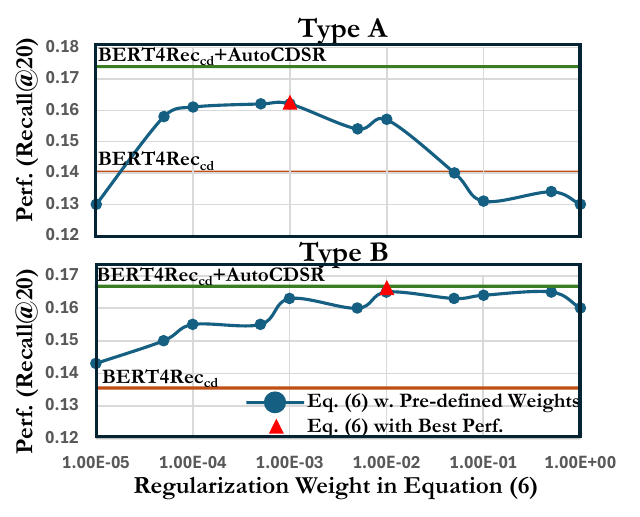}
    \vspace{-0.3in}
    \caption{
    The performance of base BERT4Rec model supervised by the additional attention loss 
    with different weights.
    Base transformers are sensitive to the weight selection and there does not exist a single optimal value for all domains. 
    }
    \label{fig:hyper_param}
    \vspace{-0.1in}
\end{figure}

\subsection{Analysis of Attention in \method}
As \method automates the knowledge transfer between different domains, we analyze the cross-domain attention scores for different stratas of samples in \cref{fig:attn_score_exp}.
Similar to settings in \cref{fig:attention_score}, `Both Correct' and `Single-domain $\checkmark$, Cross-domain \ding{55}' columns are stratas where single domain knowledge is sufficient for correct predictions and incorporating cross-domain information might introduce negative transfer.  
In these two columns, the cross-domain attention scores are significantly reduced (i.e., -17\% on the latter case), indicating that \method successfully reduces attention scores for samples where the model suffers from negative transfer.
In the meantime, for samples where knowledge transfer between domains is beneficial, the cross-domain attention score is slightly decreased (i.e., -3\%). However, this decrease is not significant compared to cases of negative transfer. 
To answer RQ3, \textbf{\method improves the CDSR performance of base transformers by selectively preserving useful cross-domain interactions while mitigating harmful ones.}

Furthermore, as shown in \cref{fig:hyper_param}, we train multiple base transformers with different weights on the additional attention loss.
We observe that although pre-defined weight assignments enhance base models, the optimal configuration varies across domains. This variation arises because such a fixed learning pattern can't effectively facilitate knowledge transfer in all cases, given the asymmetric distribution of domains across different sequences and the unintended noise introduced by additional domains.

\subsection{Computational Overhead Analysis}

\label{sec:eff}
\begin{table}
    \centering
    \vspace{-0.1in}
    \caption{Computational overheads (iterations per second, lower the better) of baselines enhanced by \method as well as state-of-the-art on KuaiRand-1K.}
    \vspace{-0.1in}
    \begin{tabular}{l|cc}
    \toprule 
    Model & Iter./s & Additional overheads $\uparrow$  (\%)\\ 
    \midrule
    BERT4Rec &  10.27 & - \\
    \; +\method   & 9.31 & 9.34 \% \\ 
    \; +\methodib & 8.22 & 19.96 \% \\ 
    SyNCRec & 2.41 & 75.53\% \\
    \bottomrule
    \end{tabular}
    \vspace{-0.2in}
    \label{tab:compute}
\end{table}
We analyze computational overheads entailed by the introduction of \method and \methodib. 
For our proposed methods, we set the number of iterations in \cref{eq:pareto} to 100 and exclude the parameters in the embedding table during the calculation of Pareto-optimal solution, as we observe that such simplifications reduce computational overheads with little to no performance degradation.
From \cref{tab:compute}, we observe that \method and \methodib introduce little computational overheads to the baseline model, making them lightweight plug-and-play extensions with little additional costs, compared to the state-of-the-art CDSR model.
To answer RQ2, this study demonstrates that the performance boost achieved by \method comes with minimal additional computational overhead, ensuring an efficient trade-off between improved effectiveness and resource consumption. Additional discussion is in \cref{app:eff}
\subsection{Robustness of \method}
\label{sec:robust}
\begin{table}
    \centering
    \caption{Recall@20 of BERT4Rec$_\text{cd}$ enhanced by our proposed methods given different corruption rate in Kuairand-1K.}
    \vspace{-0.1in}
    \begin{tabular}{l|cc|cc}
    \toprule 
    \multirow{2}{*}{Rate} & \multicolumn{2}{c}{\method} & \multicolumn{2}{c}{\methodib}\\ 
    \cmidrule(r){2-3} \cmidrule(l){4-5}
    & Type A & Type B & Type A & Type B \\
    \midrule
    0\% &  0.124 & 0.153 &  0.127 & 0.153\\
    10\% &  0.121 & 0.152 &  0.120 & 0.147 \\
    25\% &  0.122 & 0.152 &  0.117 & 0.142\\
    50\% &  0.122 & 0.151 &  0.114 & 0.141\\
    \bottomrule
    \end{tabular}
    \label{tab:corruption}
\end{table}
We analyze the robustness of \method by randomly treating a partition of behaviors in single-domain sequences as cross-domain behaviors during the training and examine if \method can recover the performance of the base transformer without this corruption. 
As shown in \cref{tab:corruption}, \method is robust to the noise in the domain knowledge and randomly mislabeling domain knowledge won't affect its performance a lot. 
However, \methodib is more sensitive to domain mislabeling, which affects the ordering of items. While \methodib excels when domain is accurate, its performance is dependent on the reliability of the domain information. 

\begin{figure}[t]
    \centering
    \vspace{-0.1in}
    \includegraphics[width=0.5\textwidth]{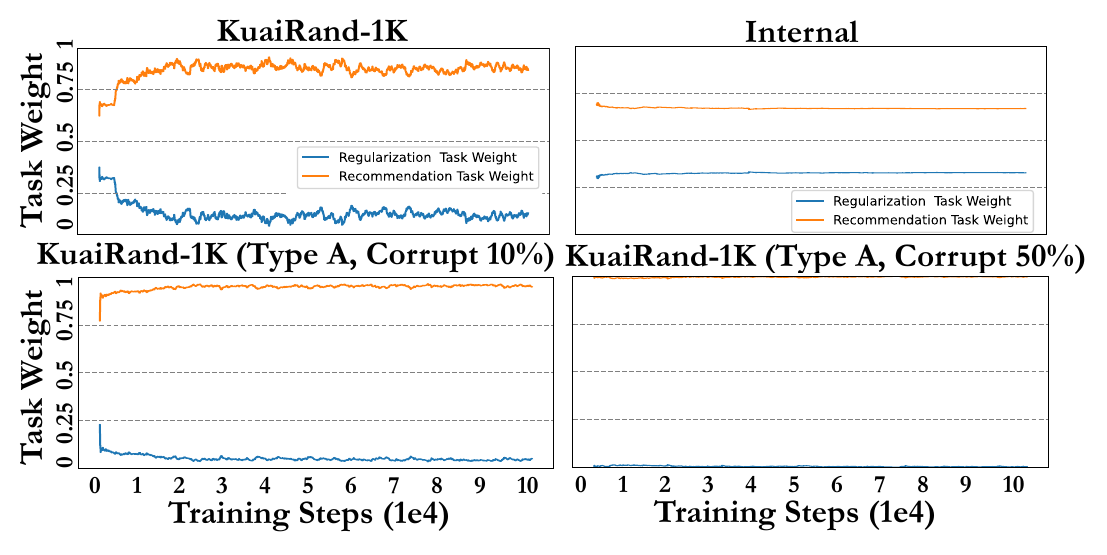}
    \vspace{-0.3in}
    \caption{
    Task weight trajectory derived by \method. 
    }
    \label{fig:task_weight}
\end{figure}

We analyze the trajectory of task weights derived by \method (\cref{fig:task_weight}). In \textsc{KuaiRand-1K}, \method gradually reduces the attention regularization weight from ~0.1 to ~0 as the corruption rate increases from 0\% to 50\%, aligning with our interpretation that cross-domain attention controls knowledge transfer. The lower regularization weight in \textsc{KuaiRand-1K} compared to \textsc{Internal} suggests greater complementarity between domains. This highlights that no static learning pattern can universally handle knowledge transfer due to asymmetric domain distributions and noise.
\section{Conclusion}
We address the challenge of knowledge transfer in CDSR by introducing \method and its enhanced variant, \methodib. Unlike existing approaches that rely on additional domain-specific components, our method improves CDSR by optimizing self-attention through a Pareto-optimal formulation. This approach dynamically balances knowledge transfer, mitigating negative transfer from noisy domains while encouraging beneficial information exchange. Extensive experiments on both large-scale production data and academic benchmarks demonstrate that \method significantly enhances transformer-based SR, achieving state-of-the-art performance with lower cost. Our work offers a fresh perspective on CDSR by leveraging self-attention to refine cross-domain learning.
Limitation of this work is discussed in \cref{app:limit}.

\bibliographystyle{ACM-Reference-Format}
\bibliography{sample-base}

\newpage
\appendix
\begin{figure*}[h]
    \centering
    \includegraphics[width=1\textwidth]{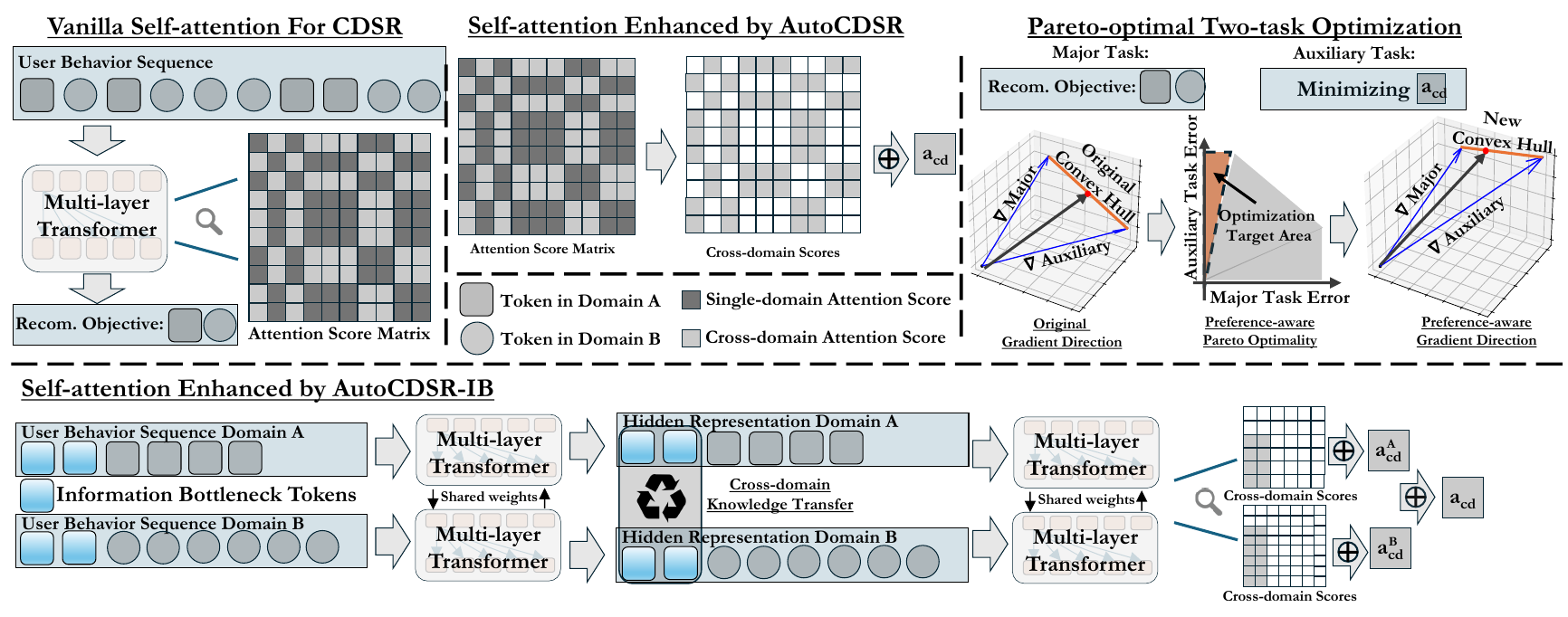}
    \caption{
    A systematic overview of our proposal. 
    (\underline{Top Left}): Training a vanilla transformer-based SR (e.g., SASRec or BERT4Rec) using user sequences containing items from multiple domains. Attention score matrix contains both single-domain attention scores as well as cross-domain attention scores.
    (\underline{Top Middle}): In \method, besides the regular recommendation objective that has been explored in SR, we extract the cross-domain attention scores and dynamically minimizes cross-domain attention scores. 
    (\underline{Top Right}): By learning a preference-aware Pareto-optimal self-attention, we reconcile between the recommendation loss $\mathcal{L}_{\text{rec}}$ and the cross-domain attention loss $\mathcal{L}_{\text{cd-attn}}$, \emph{ensuring that the self-attention increases cross-domain scores only when cross-domain information improves the recommendation task}.
    (\underline{Bottom}): Unlike \method, which permits direct information flow across domains, \methodib enforces structured information transfer by designating a small set of dedicated Pareto-optimal tokens as attention bottlenecks.
    }
    \label{fig:overview}
\end{figure*}

\newpage
\section{Appendix}
\label{sec:app}
\subsection{Additional Details on Experimental Setup}
\label{app:dataset}
\begin{table}
    \centering
    \caption{Dataset Statistics. Due to privacy constrains, we only report approximated values for \texttt{Internal} dataset.}
    \begin{tabular}{l|cccc}
    \toprule 
    Dataset & \# Users & \# Items & \# Interactions & Sparsity \\
    \cmidrule(r){1-5} 
    \multicolumn{5}{c}{\texttt{Amazon-Review}} \\
    \cmidrule(r){1-5} 
     \texttt{Book}  & \multirow{5}{*}{105,364}  & 425,985  & 1,422,676 & 99.98\% \\
     \texttt{Clothing}  &   & 290,804  & 947,417 & 99.98\% \\
     \texttt{Video}  &   & 29,013  & 292,891 & 99.97\% \\
     \texttt{Toys}  &   & 121,559  & 575,499 & 99.98\% \\
     \texttt{Sports}  &   & 133,066  & 541,717 & 99.99\% \\
     \cmidrule(r){1-5} 
    \multicolumn{5}{c}{\texttt{KuaiRand-1K}} \\
    \cmidrule(r){1-5} 
     \texttt{Type A} & \multirow{2}{*}{5,101}  & 1,435,814  & 5,014,567 & 99.93\%  \\
     \texttt{Type B} &   & 325,734  & 967,784 & 99.94\%  \\
     \cmidrule(r){1-5} 
     \multicolumn{5}{c}{\texttt{Internal}} \\
     \cmidrule(r){1-5} 
     \texttt{Tab A} & \multirow{2}{*}{$\sim$5M} & 
     $\sim$12M & $\sim$900M & 99.99\% \\
     \texttt{Tab B} &  & 
     $\sim$3M & $\sim$170M & 99.99\% \\ 
    \bottomrule
    \end{tabular}
    \label{tab:dataset}
\end{table}
\noindent \textbf{Datasets}:
We conduct experiments on two academic benchmark datasets (i.e., \textsc{KuaiRand-1K}~\citep{gao2022kuairand} and \textsc{Amazon Reviews}~\citep{mcauley2015image,hou2024bridging}) as well as one industrial dataset collected internally at a leading social media platform spanning interactions over multiple mobile in-app surfaces (dubbed as \textsc{Internal}).
For \textsc{KuaiRand-1K}, we use music type as the domain indicator. Specifically, we treat items with music type "9" as Type A and "4" as Type B, two most frequent music types in this dataset, and treat all other items as an unique category. Since the sequence length distribution in this dataset is very skewed, we split long user sequences into multiple subsequences and treat different subsequences as unique users, where each subsequence could have a maximum amount of 800 interactions in all domains, leading to 5,101 total number of users.
For \textsc{Amazon Reviews}, following the same setting explored in the previous work~\citep{park2024pacer}, we include user behaviors from Book, Clothing, Video, Toy, and Sports domains. 
Whereas for \textsc{Internal}, we use the in-app surface where the user behavior comes from as the domain indicator. 
Due to legal constraints, we are not allowed to disclose further detailed information about \textsc{Internal} dataset.
Statistics of these processed datasets are shown in \cref{tab:dataset}.

\vspace{0.1in}
\noindent \textbf{Baselines}:
We compare \method and \methodib against a broad range of baselines, including domain-agnostic sequential recommenders (i.e., GRU4Rec~\citep{hidasi2015session}, SASRec~\citep{kang2018self}, BERT4Rec~\citep{sun2019bert4rec}) and domain-aware sequential recommenders (i.e., SyNCRec~\citep{park2024pacer}, C$^2$DSR~\citep{cao2022contrastive}, and CGRec~\citep{cao2022contrastive}). 
As \method and \methodib improve the self-attention mechanism in the transformer model, we utilize them as plug-and-play modules and apply them to vanilla transformer-based CDSR models (i.e., SASRec and BERT4Rec). 
For \textsc{Amazon Reviews}, we directly re-use the baseline performance reported in the previous work~\citep{park2024pacer} and follow the same setting as this paper explores. 

\vspace{0.1in}
\noindent \textbf{Evaluation Protocol}:
We evaluate with metrics adopted in previous works, including NDCG, Recall, and Hits~\citep{park2024pacer,kang2018self}. All studies are repeated for 5 times and mean values are reported.
We explore the "leave-one-out" data split strategy~\citep{kang2018self}, where the last item in the user sequence is used for testing, the second last is used for validation, and all remaining items are used for training.
We adopt an early stopping strategy, where the training will be terminated if the validation NDCG stops increasing for 2,000 continuous training steps and we check the validation metrics every 500 steps. 
We use models with the best validation performance to report the performance. 
Besides, the evaluation metrics are computed against a batch of random negatives in the whole item candidate pool to reduce the evaluation overheads and we use 100, 50,000 and 500,000 random negatives for \textsc{Amazon-Review}, \textsc{KuaiRand-1K}, and \textsc{Internal} respectively.

\vspace{0.1in}
\noindent \textbf{Hyper-parameter Tuning}:
For baseline implementation and hyper-parameter tuning for baseline models, we tune the learning rate and weight decay both within the range of \{5e${^{-3}}$, 1e${^{-3}}$, 5e${^{-4}}$, 1e${^{-4}}$\}. 
We randomly choose 5 combinations for each model and report the performance of the best combination. 
Furthermore, to ensure a fair comparison, we set the embedding dimensions for all models to 32 (i.e., $d=32$) and use the same amount of transformer layers and heads for transformer-based models (i.e., 6 layers with 4 heads in each layer).
We explore AdamW optimizer to train our model and a cosine warm-up strategy with 1,000 warm-up steps.
We use a batch size of 32, 128 and 128 for \textsc{Internal}, \textsc{KuaiRand-1K}, and \textsc{Amazon-Review} respectively.

\vspace{0.1in}
\noindent \textbf{Software and Hardware}:
We explore HuggingFace\footnote{https://huggingface.co/docs/transformers/en/index} to implement all base transformers and Lightning\footnote{https://lightning.ai/docs/pytorch/stable/} to build our training infrastructure.
For training, we use n1-standard-96 (96 vCPUs, 360 GB Memory) machines on Google Could Platform with 8 V100s with 16 GB VRAM each to train all of our models.
\subsection{Discussion w.r.t. Efficiency of \methodib}
\label{app:eff}
In \cref{tab:compute}, \methodib runs slower than \method, which may seem counterintuitive since \methodib decomposes the full attention calculation into multiple smaller attention calculations within different domains. This behavior arises from our current implementation, where we use attention masks to simulate domain-specific attention. As a result, each single-domain sequence retains the same length as a cross-domain sequence, with cross-domain items simply masked out.
We adopted this approach for ease of development and faster model iteration. However, \methodib incurs additional computational overhead by performing attention calculations on masked-out cross-domain items, even though these computations are ultimately discarded. In practice, this inefficiency could be mitigated by developing a more advanced data preprocessing pipeline, allowing \methodib to focus only on the necessary computations and eliminate wasteful processing.

\subsection{Sensitivity to the Number of Iterations in Deriving the Pareto-optimal Solution}
\label{app:sensitivity}
\begin{table}
    \centering
    \vspace{-0.1in}
    \caption{The performance (i.e., Recall@20) and training throughput (i.e., the number of training iterations per second) of \method and \methodib on Type A of \textsc{KuaiRand-1K} with different maximum number of iterations on deriving the Pareto-optimal solution.
    Backbone model used is BERT4Rec}
    \vspace{-0.1in}
    \begin{tabular}{l|cccc}
    \toprule 
    \multirow{2}{*}{\# of Iter.} & \multicolumn{2}{c}{\method} & \multicolumn{2}{c}{\methodib}\\ 
    \cmidrule(r){2-3} \cmidrule(l){4-5}
    & Perf.  & Throughput & Perf.  & Throughput \\
    \midrule
    10 &  0.170 & 9.66 &  0.134 & 8.57\\
    100 &  0.173 & 9.31 &  0.139 & 8.22 \\
    200 &  0.173 & 9.30 &  0.139 & 8.13\\
    500 &  0.173 & 9.30 &  0.139 & 8.12\\
    \bottomrule
    \end{tabular}
    \label{tab:iteration}
\end{table}
We analyze the sensitivity of \method and \methodib to the maximum number of iterations in deriving the Pareto-optimal solution, as shown in \cref{tab:iteration}. Our results indicate that the training throughput of both \method and \methodib saturates at 100 iterations, with no further improvements beyond this point. This is because the Pareto-optimal solution converges within 100 iterations, triggering an early stopping mechanism and preventing unnecessary computations.

\subsection{Experimental Setup in \cref{tab:intro_motivtion}}
\label{app:intro_setup}
In this study, we train multiple BERT4Rec models using different input sequences. For single-domain models (column (i)), we exclude items outside the respective domain and train a separate BERT4Rec model for each domain.
For cross-domain models (column (ii)), we incorporate cross-domain items into the input sequence up to the prediction target used in the single-domain model. This ensures a fair comparison, as each domain still has its own BERT4Rec model, but with enriched input sequences that include cross-domain context.

\subsection{Limitation}
\label{app:limit}
As mentioned in \cref{sec:robust}, one limitation of \methodib could be its high sensitivity to noise in the domain labeling.
While \methodib excels when domain is accurate, its performance is dependent on the reliability of the domain information. 
Hence when the domain information is unknown or noisy, we recommend using \method instead. 
Furthermore, we observe no ethical concern entailed by our proposal, but we note that both ethical or unethical applications based on sequential recommendation may benefit from the effectiveness of our work. 
Care should be taken to ensure socially positive and beneficial results of machine learning algorithms.
\newpage
\end{document}